\documentclass[aps,prl,showkeys,superscriptaddress,10pt,nobibnotes,twocolumn]{revtex4-1} 
\usepackage{soul}
\usepackage{lipsum}
\usepackage{amsmath,bm}
\usepackage{graphicx}
\usepackage[dvipsnames]{xcolor}
\usepackage{cancel}

\newcommand{\lw}{\linewidth}
\graphicspath{{./fig/}}

\begin{document}
\title{Topology controls magnetoelectric switching in spiral multiferroics}

\author{Francesco Foggetti}
\affiliation{Quantum Materials Theory, Italian Institute of Technology, Via Morego 30, 16163 Genova, Italy}
\affiliation{Department of Physics, University of Genova, Via Dodecaneso, 33, 16146 Genova GE}
\affiliation{Department of Physics and Astronomy, Uppsala University, P.O. Box 516, S-751 20 Uppsala, Sweden}

\author{Margherita Parodi}
\affiliation{Quantum Materials Theory, Italian Institute of Technology, Via Morego 30, 16163 Genova, Italy}
\affiliation{Department of Physics, University of Genova, Via Dodecaneso, 33, 16146 Genova GE}

\author{Naoto Nagaosa}
\affiliation{RIKEN Center for Emergent Matter Science (CEMS), Wako, Saitama 351-0198, Japan} 

\author{Sergey Artyukhin}
\affiliation{Quantum Materials Theory, Italian Institute of Technology, Via Morego 30, 16163 Genova, Italy}


\begin{abstract}

Magnetoelectric multiferroics, hosting both magnetism and ferroelectric
polarization, enable switching of the magnetic order by electric fields,
which is vital for spintronics and information storage devices. Spiral
multiferroics have long held a promise for electric control of
magnetization, although how the magnetic switching occurs under an
applied electric field remains enigmatic. We find that, in contrast to
all known bulk ferroics, where switching dynamics is confined to domain
walls, in spiral multiferroics it can have nonlocal nature. In materials
where domain walls have non-trivial topology, such dynamics does not
only require the rotation of spins within walls but also in an entire
adjacent domain. This unprecedented nonlocal dynamics, where
the outermost wall accelerates as it gets closer to the surface, gives
favorable scaling for miniature devices. Inner walls do not move at all,
giving vanishing dielectric response. Our findings establish a paradigm
of nonlocal switching dynamics in ferroic materials.
%
%

\end{abstract}
\maketitle

{\bf Introduction --- }Competing spin interactions can lead to non-collinear spin textures, such as spin spirals and skyrmions \cite{Diep1994book}. The ability to manipulate magnetic orders electrically is vital for technology. 
%
%
The discovery of ferroelectric polarization induced in spiral magnets \cite{Kimura2003} fueled the explosion of interest in multiferroics and non-collinear magnets over the past 20 years \cite{Karna2021,Cohen2020,Qaiumzadeh2018,Schoenherr2018}, due to promise of electric control of  magnetism, however switching in spiral magnets themselves is still poorly understood \cite{Kagawa2009,McQuaid2017,Brierley2014,Whyte2015,dong2019magnetoelectricity}. 

Here we find that electric field can induce unconventional nonlocal dynamics of domain walls (DWs) in spiral magnets. There are two principal types of low-energy walls in cycloidal spiral magnets (see Fig.~\ref{fig:type12}). The walls that have coplanar spins (type I) exhibit nonlocal motion,
{\it i.e.} spins far from the wall rotate. The motion of type II walls only requires spins within the walls to rotate, similarly to ferromagnetic DWs. Nonlocal dynamics of type I walls leads to unconventional equations of motion, zero contribution to dielectric constant in thermodynamic limit and a much stronger pinning compared to type II walls, as well as a peculiar breakdown at high electric fields. The results reveal an unprecedented switching behavior and emphasize the importance of the topology of the ground state (encoded in the sense of spin rotation that cannot be changed locally) for the dynamics in the most basic non-collinear magnets.

{\bf Model --- }
A spin spiral state is usually stabilized by competing exchange interactions, as in $R$MnO$_3$, YMn$_2$O$_5$, MnWO$_4$ \cite{Kimura2003,Cheong2007} and CuO-based materials that support spiral order up to room temperature \cite{Katsura2005,Rocquefelte2013,Terada2022}. A cycloidal spiral spin configuration, where the spin rotation plane contains the wave vector, breaks inversion symmetry, and inverse Dzyaloshinskii-Moriya (DM) effect leads to bond bending and a ferroelectric polarization $\mathbf{P}$, as shown in Fig.~\ref{fig:Ansatz} \cite{Katsura2005,Mostovoy2008,Sergienko2006}. The two domains with opposite spin rotation sense (or opposite chirality) are separated by a \textit{chiral} domain wall \cite{Li2019chiral}. In this context ``chiral" refers to a system which is not identical to its mirror image.
The domain wall is simultaneously magnetic and ferroelectric, therefore such a {\it multiferroic domain wall} can be driven by an electric field and its motion induces simultaneous changes in both ferroelectric and magnetic orders.

%



{

\begin{figure*}[t]
    \centering
    \includegraphics[width=\linewidth]{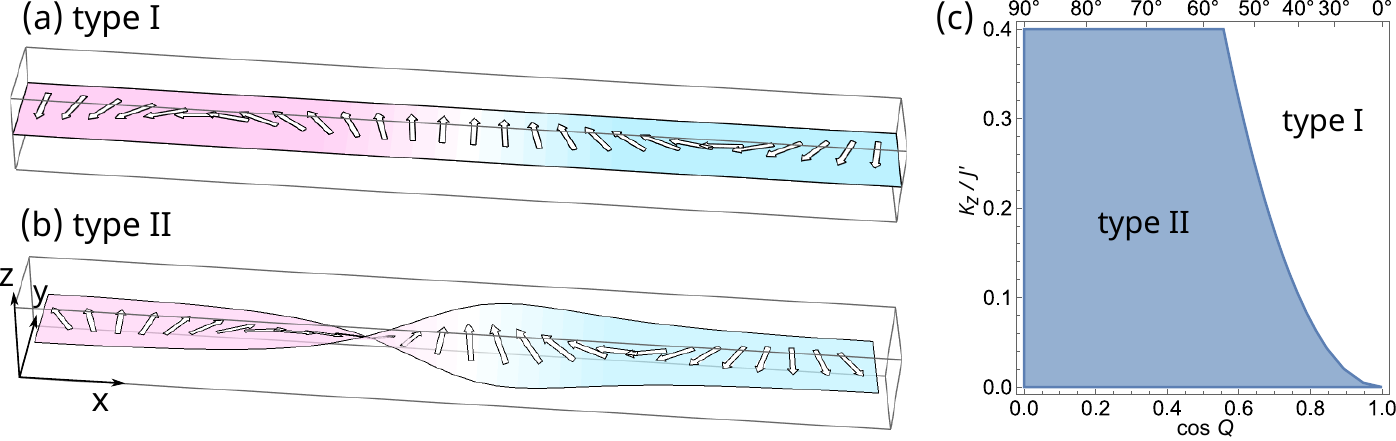}
    \caption{{\bf Two types of ferroelectric domain walls in spiral multiferroics.} (a) type I wall where the direction of the spin rotation is reversed while spins always remain within the fixed spin plane. The winding of spins around the $z$ axis gives rise to non-trivial topology. (b) type II wall with the gradual twisting of the spin rotation plane. The winding number is ill-defined since the plane passes through the $z$ axis. (c) Phase diagram for the relative stability of the two types of the domain walls in the plane of spin anisotropy $K_z/J_2$ and $\cos Q$ ($Q$: the wavenumber of the spiral). The numbers at the top of the frame show the corresponding spin rotation angle between the cells in degrees, $Q\cdot 180^\circ /\pi$.}
    \label{fig:type12}
\end{figure*}

}

{
In order to study the motion of the chiral DWs, we consider a 1D model
, where a magnetization $\mathbf{M}(x)$ continuously varies along $x$ through an interval of total length $L$. 
The dynamics of the spiral magnet is described by the Lagrangian density $\mathcal{L}$ \cite{Li2012,auerbach2012book},
\begin{equation}                                                                                          %
\begin{split}                                                                                             %
    \mathcal{L}=&\alpha \mathbf{A}(\mathbf{M})\cdot                                                       %
    \dot{\mathbf{M}}-J(\nabla\mathbf{M})^2-J'(\nabla^2\mathbf{M})^2 +\\&+\gamma \mathbf{[\mathbf{E}\times %
    \mathbf{x}]}\cdot[\mathbf{M}\times\nabla\mathbf{M}]-K_z M_z^2,                                        %
\label{eq:Lag}                                                                                            %
\end{split}                                                                                               %
\end{equation}                                                                                            %
where the Weiss-Zumino-Witten term with $\alpha=\hbar/M^2$ results in the precession of the magnetization, with $\mathbf{A}$ acting as a magnetic monopole vector potential, so that $\nabla_\mathbf{M} \times\mathbf{A}(\mathbf{M})=\mathbf{M}$ \cite{auerbach2012book}. Competing magnetic exchange interactions $J<0$ and $J'>0$ produce a magnetic spiral along $x$ with the wavevector $Q=\sqrt{-J/2J'}$. $\mathbf{E}$ is the electric field parallel to $\mathbf{y}$, and $\gamma$ represents the DM interaction and is proportional to the spin-orbit coupling constant. $K_z$ is the hard axis anisotropy constant that forces spins into the $xy$-plane. 
%
We describe the energy dissipation using the Rayleigh dissipation functional density
 $\mathcal{R}= \beta\dot{\mathbf{M}}^2$
 , with the damping constant $\beta$ having a dimension of a momentum. 
In numerical simulations we use a discrete model with competing nearest-neighbor ($J_1$) and next-nearest neighbor ($J_2$) interactions (see Supplementary for details).

{\bf Types of walls --- } Depending on the parameters of the model and the orientation of the DW with respect to the wave vector, walls of multiple types exist \cite{Li2012}. Walls whose plane contains the spiral wave vector are formed by an array of vortices or merons, and have a high energy \cite{Li2012}. Experimental observations suggest that such walls disappear quickly \cite{Biesenkamp2021}. The walls whose plane is perpendicular to the wave vector, so called Hubert walls \cite{Hubert1974theorie}, come in two types. In type-I DW, spins change their rotation sense while remaining within the same plane. Such walls are narrow, having the width of the order of the spiral period, and are stabilized by large easy-axis anisotropy. In type-II walls, spin rotation plane is twisting continuously, so that spins leave the easy plane within the wall. Figure 1(c) shows the regions of model parameters that correspond to type-I and type-II walls. See the Supplementary Material for the energetics of type I and type II domain walls which determines the phase diagram Fig.~\ref{fig:type12}(c). The important distinction between type-I and type-II walls is their topology. In type-I walls, all spins lie in the same plane, and their polar angle $\phi(x)$ is changing in space {\it continuously},  cf. Fig.~\ref{fig:Ansatz}(e), as opposed to type-II DWs, where spin rotation plane rotates and passes through the vertical axis, where $\phi$ experiences jumps. A single type-I wall inside the bulk cannot move without affecting the neighbors, and the ferroelectric polarization, defined by the winding number $\int [M\times \nabla M]_z dx =\phi(L)-\phi(0)$, is not affected by the motion of pairs of inner walls. Hence, the electric field only applies pressure to the outermost DWs whose motion alters $\phi(0)$ and $\phi(L)$. It is this difference in topology that results in a profound difference in physical properties between type I and type II walls.

{\bf Type-I walls --- } We now begin with type-I walls that cannot be moved without rotating spins in the domains adjacent to the wall. The required spin rotation in the whole domain is shown by the dashed line in Fig.~\ref{fig:Ansatz}(d), where spin angles before (after) the rotation are indicated by solid (dashed) lines. 

Experimental measurements \cite{Biesenkamp2021} and references therein suggest that the domains take the form of relatively straight stripes perpendicular to the spiral wave vector. For such a quasi-1D domain configuration with DWs at positions $X_i$, $i=1\dots N$, shown in Fig.~\ref{fig:Ansatz}(e), we express the magnetization in spherical coordinates, $\mathbf{M}=(\sin\theta\cos\phi,\sin\theta\sin\phi,\cos\theta)$, where the spatial dependence of the angles $\theta$ and $\phi$ is given by the following Ansatz that captures the essential physics of the DW motion: $\phi(x)=\phi_0-Qx+{2}Q(x-X_0)\Theta(x-X_0)$ and $\cos\theta=\delta_1$ for $x<X_0$ and $\cos\theta=\delta_2$ otherwise, as illustrated in Fig.~\ref{fig:Ansatz}(d).
Here $\phi_0$ encodes the spin orientation at $x=0$, 
$\Theta$ is the Heaviside step function and $\delta_{1,2}$ (with $\delta_{1,2}\ll 1$) represent the out-of-plane components of the spins. 
The presence of the $M_z$ components of the magnetization encoded by $\delta_{1,2}$ is needed for the magnetization to rotate in $xy$ plane. Indeed, for the DW to move, spins must precess around $z$ axis. Magnetization precesses around the effective field $\mathbf{h}=-\partial \mathcal H/\partial \mathbf{M}$, 
that has no $z$ component when the spins lie in the $xy$-plane. 
The precession around the $z$ axis becomes possible because the magnetization acquires $M_z$ component due to DM field, that has a component in $xy$ plane perpendicular to the magnetization. A variational Ansatz for $N$ walls at positions $X_0,\dots,X_N$ is shown in Fig.~\ref{fig:Ansatz}(e). Variation of $\phi_i$, corresponding to a phason mode, rotates all spins in the $i$-th domain. The positions of the walls are given by intersections of the straight segments, $\phi_i(X_i)=\phi_{i+1}(X_i)$ as $X_i=(-1)^i(\phi_{i+1}-\phi_i)/(2Q)$.
}

\begin{figure}[t]                                                                           %
    \centering                                                                              %
    \includegraphics[width=\linewidth]{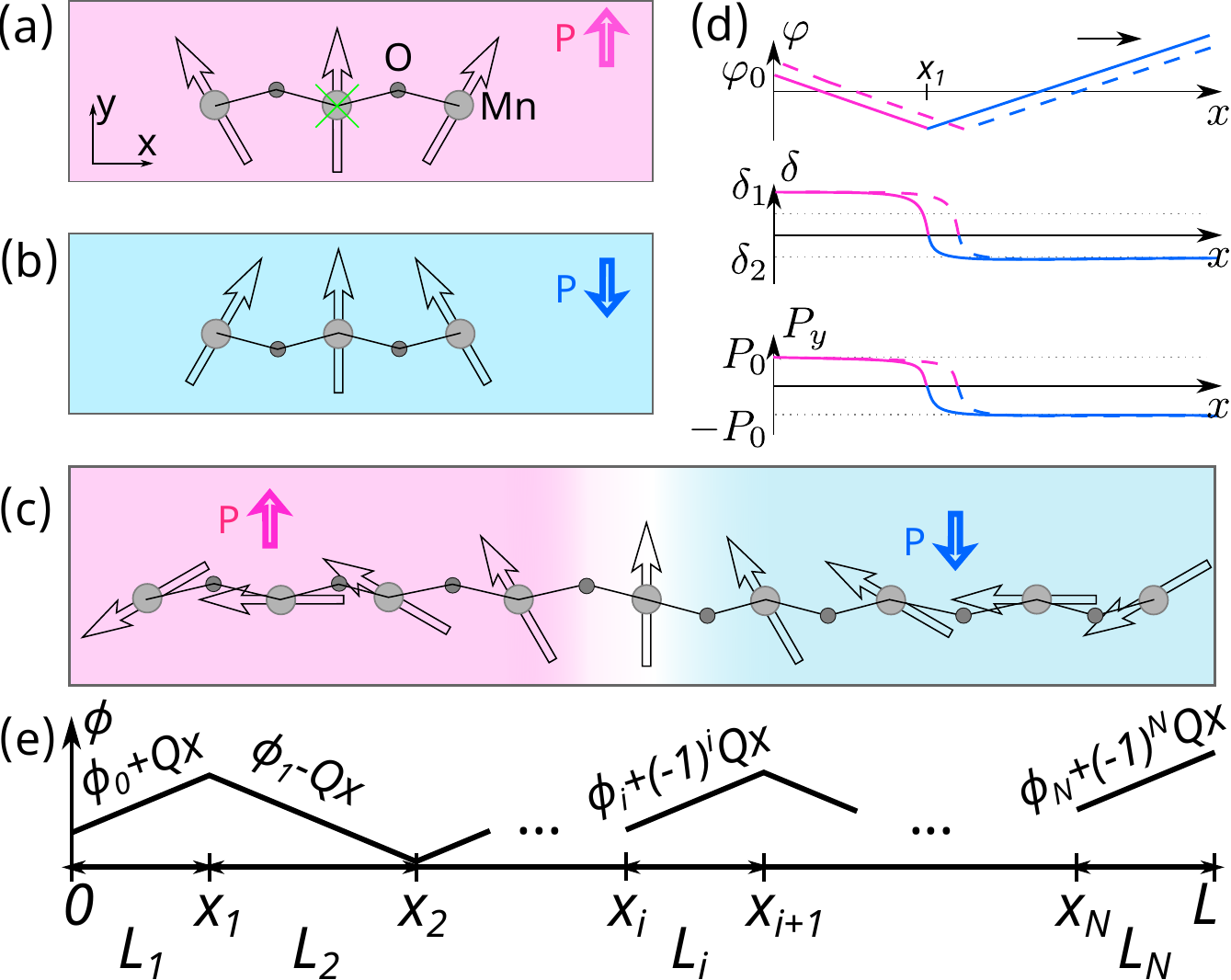}    
    \caption{\label{fig:Ansatz} {\bf Type-I multiferroic domain walls.} (a) Cycloidal spiral domain, corresponding to a specific ferroelectric polarization $\mathbf{P}$ of the sample, and (b) a domain, obtained by inversion, where the spin rotation sense is reversed as is the ferroelectric polarization. The rotation sense of spins defines the sign of the ferroelectric polarization. (c) Cycloidal spiral texture with a domain wall at the center. The spin rotation sense and the ferroelectric polarization change at the wall. Colors correspond to the direction of $\mathbf{P}$. (d) The corresponding profiles of the polar angle $\phi(x)$, the out of plane component of the magnetization $M_z(x)=\delta=\cos \theta(x)$ and ferroelectric polarization component $P_y(x)$. (e) Multi-domain case, the spin-rotation sense changes at every wall.
   }
\end{figure}
%

\begin{figure*}[t]
    \centering
    \includegraphics[width=\lw]{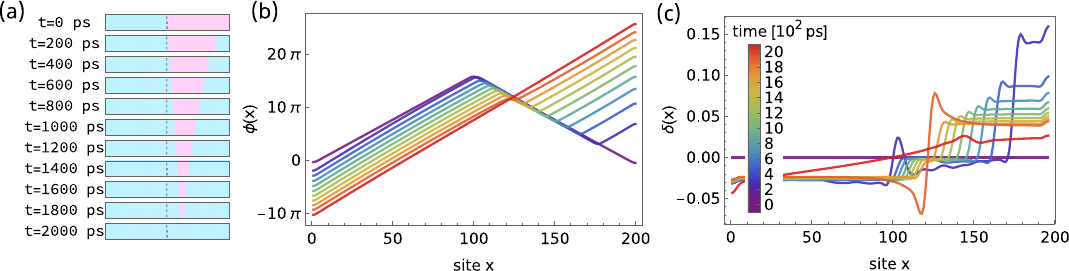}
    \caption{\label{fig:simul} {\bf $E$-field driven motion of type-I walls} (a) Temporal evolution of a polarization profile ($+P_y$ in pink, $-P_y$ in blue, time indicated in ps). A chiral DW is initially placed at the center (marked with a dashed black line), and another one is nucleated at the right boundary.
    (b) Profiles of $\phi(r)$, the polar angle characterizing the spin direction. Color encodes different times, indicated by the color scale in panel (c). The linear behaviour $\phi(r)\approx Q r$ corresponds to a spiral domain, and the V-shaped cusps correspond to chiral (and ferroelectric) DWs. At $t=0$ a single wall is present at the center. The slopes for $t=0$ and at later times are slightly different because the wave vector is adjusted by the $\mathbf{E}$ field via the DM interaction. (c) The spatial profiles of the out-of-plane spin component, $\delta(x)$, 
    obtained from the LLG simulations. 
    $\delta(x)$ satisfies $\int dx \delta(x)=0$.}
\end{figure*}

In the spirit of a variational approach, we integrate $\mathcal{L}$ and  $\mathcal{R}$ over each domain to obtain 
\begin{equation}
    \mathcal{L}=\sum_i^N\mathcal{L}_i{+\mathcal{L}_E}, \quad\mathcal{L}_i=\left[\alpha\dot\phi_i\delta_i+\frac{1}{2m}\delta_i^2\right]L_i^{(0)},
\end{equation}
with $m^{-1}=2Q^2(J+Q^2J')+K_z$. Hence, all $\phi_i$'s, representing phason modes, or uniform rotations of all spins in each domain, are decoupled and evolve independently. The integrals are proportional to the instantaneous domain widths $L_i^{(0)}$, much larger than the contribution of DWs, proportional to the wall width, which justifies the Ansatz with sharp walls. The out-of-plane spin component plays the role of the canonical momentum $p_{\phi_i}=\partial \mathcal{L}/\partial{\dot{\phi}_i}=\alpha\delta_i$, which enters quadratically, giving rise to inertia. Remarkably, the ferroelectric polarization is determined by the spin winding number, hence the interaction with the electric field only involves the phases of the outermost domains, 
$\mathcal{L}_E=\gamma E_y(\phi_0-\phi_N)$, so only the walls closest to the surface are driven by the $\mathbf{E}$ field. This means that the type-I domain walls do not contribute to the dielectric response in the thermodynamic limit, in sharp contrast to the type-II walls. Note that we neglect $\delta^2$ in the equations of motion and therefore this result is valid only in the linear response regime.

For a single wall, using Euler-Lagrange equations for $X_1,\phi_0,\delta_1$, and $\delta_2$ (details in the Methods section), equivalent to Thiele equations with the DW coordinate and out-of-plane component as effective coordinates, we obtain the steady state velocity
\begin{eqnarray}
    \dot{X_1}&=&-\frac{\gamma E}{4\beta Q}\frac{L}{X_1(L-X_1)},
    \label{eq:X1}
\end{eqnarray}
to a linear order in $E$ \cite{NoteM}. It is remarkable that the DW velocity is inversely proportional to the distances of the DW from the sample boundaries, $X$ and $L-X$. 
For $N\geq2$ walls shown in Fig.~\ref{fig:Ansatz}(e) we obtain
\begin{equation}
    \dot X_1=-\frac{\gamma E}{4Q\beta L_1},\;
    \dot X_N=\frac{\gamma E}{4Q\beta L_N}.
    \label{eq:X2}
\end{equation}
Unlike in ferro- and antiferromagnets, the speeds of the walls,  are not only determined by the field but by the domain structure. Remarkably, the inner walls do not move.

Spin precession around $z$ axis, necessary for the DW motion, is driven by an out-of-plane components of magnetization  $\delta_{1,2}$, induced by the electric field. We find them to be inversely proportional to the outer domain width, $\delta_{0,N}\propto E/L_{0,N}^{(0)}$, such that
$\delta_1L_0+\delta_NL_N=0$ \cite{NoteM}.
The integration of Eq.~(\ref{eq:X1}) gives for $X\ll L$
\begin{equation}
    X(t)=\left(X^2(t_0)-\frac{\gamma E}{2\beta Q}(t-t_0)\right)^{1/2}.\label{eq:Xt}
\end{equation}
%
%
Close to the surface, the velocity in Eqs.~(\ref{eq:X1},\ref{eq:X2}) diverges due to our simplified Anzats with zero DW widths. As our simulations show, the wall speeds will not exceed the magnon velocity.

{\bf Type-II walls --- } The energy of type-II walls can be expressed in terms of the unit  vector ${\bf n}({\bf r})$, normal to the local spin rotation plane,
$$F=\int d{\bf r} \left( -J(\nabla {\bf n})^2-K_zn_z^2-P_0E_zn_z\right),$$
with $-J>0$ playing the role of stiffness (see SI for details).
Energy minimization leads to a sine-Gordon equation with a solution ${n_z=\tanh\left(\frac{x-x_0}{\lambda}\right)}$. The domain wall is wide, $\lambda=\sqrt{J'/K_z}\gg Q^{-1}$, and thus such walls are mobile, potentially giving rise to a large dielectric constant \cite{Kagawa2009}. It follows from Thiele equations \cite{thiele1973steady} that under an electric field $E_z$ such wall will move with a velocity $v_x=2P_0E/\alpha$, where $P_0$ is the ferroelectric polarization induced by inverse DM interaction, and $\alpha$ is the Gilbert damping. The motion of such wall only involves spin rotation within the wall.

{\bf Spin dynamics simulations --- }In order to describe the essential physics of $\mathbf{E}$-field driven DW motion, our model relied on a number of approximations, such as sharp (zero-width) DWs and piecewise-constant $\theta(x)$. Here we validate our model by performing atomistic spin dynamics simulations of a discrete $J_1$-$J_2$ model using the UppASD code \cite{Skubic2008} (see Supplementary Information for details).

The rotation of the spins can be visualized by plotting $\phi(r)$, cf. Fig.~\ref{fig:simul}(b). Straight lines $\phi(r)=\pm Q r$ represent uniform spin spiral and correspond to chiral domains, while V-shaped cusps are the DWs. Figure~\ref{fig:simul} illustrates the domain wall motion when a single domain is initially present in the middle of the sample ($t=0$~ps). That corresponds to a violet V-shaped curve for $\phi(x)$ in Fig.~\ref{fig:simul}(b) and the top polarization profile in Fig.~\ref{fig:simul}(a). At $t=200$~ps a V-shape appears near the right boundary in Fig.~\ref{fig:simul}(b), indicating a nucleation of a second domain with $P$ along the external field. The large distances between the right cusps at subsequent steps indicate large initial speed of the right DW, that is consistent with Eq.~(\ref{eq:X2}) 
and the fact that the right domain is very narrow (small $L_2$). As the nucleated domain on the right grows, the wall gets slower and the distances between the cusps decrease, as seen in Fig.~\ref{fig:simul}(a,b). The wall initially at the center starts with a lower velocity and just slightly accelerates while moving to the right.


The DW at the center is slower than the one at the right boundary, due to the greater number of spins that have to be rotated by the electric field-induced torque for the central wall to be moved, which is apparent from the spin rotation speed, indicated by the profile of the out-of-plane component in Fig.~\ref{fig:simul}(c). The two walls move towards each other, collide, and annihilate at approximately {2000}~ps, leaving the system in a single domain state. The velocities and trajectories of the walls are in good agreement with the theory, cf. Fig.~{\ref{fig:fitcompare}} \cite{NoteM}.

\begin{figure*}
    \includegraphics[width=\linewidth]{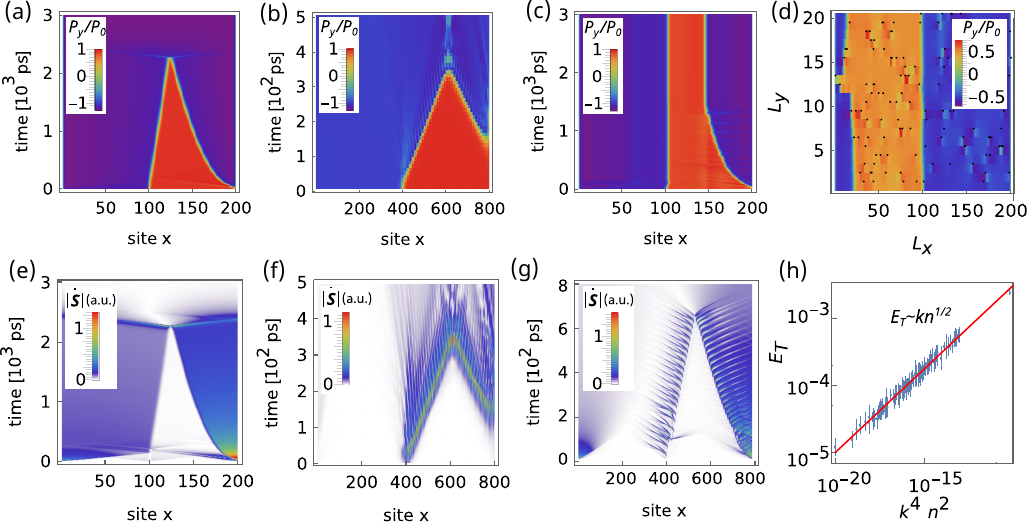}
    \caption{\label{fig:disorder}{\bf Dynamics of type-I and II walls, pinning and breakdown regime}
     (a,e) polarization $P_y(x,t)$ and spin rotation velocity $|\dot{\mathbf{S}}(x,t)|$ dependencies on the $x$ coordinate and time for switching of type-I DWs under electric field. Panel (f) indicates that spins within the entire outer domains rotate.
    (b,f) Respectively, polarization $P_y(x,t)$ and spin rotation velocity $|\dot{\mathbf{S}}(x,t)|$ dependencies on the $x$ coordinate and time for switching of type-II DWs under electric field. Panel (e) indicates that only spins within the DWs rotate.
    (c,d) Switching in the type-I system in the presence of disorder, represented by impurities with concentration $n$ inducing an easy axis anisotropy of strength $k$  in a random direction within the $xy$ plane. In Panel (c) the DW is ultimately pinned at the depth around 150th site. Panel (d) shows terrace formation on the DW pinned around $x=$180. Impurities are represented as black dots; distortions of the spiral order near impurities are seen.
    (g) Walker breakdown-like behavior in a type-I system under high electric field.
    (h) Scaling of the threshold electric field $E_T$, necessary to depin the DW at 50th site, with impurity concentration $n$ and anisotropy strength $k$.
    }
\end{figure*}

To visualize the rotation of individual spins, we plot $|\dot{\mathbf{S}}(x,t)|$ in Fig.~\ref{fig:disorder}(e,f). In contrast to 
type-II walls, where the wall motion is facilitated by rotation of spins exclusively within the DWs (panel e), for type-I system spins rotate in the entire outermost domains (panel f). Upon collision of the walls and collapse of the middle domain, magnons are excited and propagate away from the collision site, as seen in Fig.~\ref{fig:disorder}(e,f). Corresponding polarization change in Fig.~\ref{fig:disorder}(a,b) suggests these are electromagnons, i.e. carry an electric dipole moment. Magnon scattering from the edge of the system and from the walls are also apparent in the right domain at low $t$ in Fig.~\ref{fig:disorder}(b,f). 

{\bf High-field breakdown ---} In type-II DWs the $\mathbf{E}$ field through DM interaction is directly creating effective magnetic field that is responsible for the associated spin precession resulting in DW motion. At high fields the wall velocity approaches the magnon velocity, and there is no breakdown.

In contrast, in type-I DWs the $\mathbf{E}$ field competes with the easy $xy$ plane anisotropy to induce the out-of plane spin component $\delta$, resulting in spin rotation. 
This situation is analogous to Walker breakdown in ferromagnets \cite{Walker1974}. At high $E$ fields 
the torque due to the DM interaction overcomes the easy plane anisotropy, leading to spin rotations outside of the $xy$ plane, and a chaotic breakdown behavior seen in Fig.~\ref{fig:disorder}(g). Bright streaks correspond to magnons from nucleation of favored domains inside the unfavored one and their merger with the growing domain.

{\bf Pinning by disorder---} In experiments spiral magnets may be influenced by many types of disorder \cite{Nattermann2014EPL}. Here we simulate impurities placed at $\mathbf{r}_j$ that produce local anisotropy of strength $k$ along a randomly oriented easy-axis $\mathbf{n}_i$ on the neighboring spin, thus giving the energy contribution $H_i=\sum_j k (\mathbf{M}(\mathbf{r}_j)\cdot \mathbf{n}_j)^2$. The disorder of this form appears in the problem of charge density wave pinning \cite{Gruner1988RMP} and is known in the weak disorder limit to lead to the scaling of the threshold depinning field as $E_T\sim k^4 n^2$ in the collective pinning case. Hence, we performed simulations for different $k$ and $n$ and found scaling with respect to the same combination of parameters but with a different exponent, $E_T\sim (k^4 n^2)^{1/4}$, as seen in Fig.~\ref{fig:disorder}(h).

This may be due to the following reason. The competition of the exchange energy in Eq.~(\ref{eq:Lag}) and the disorder sets the length scale, below which exchange sets spin correlations, known as Larkin length $R$ \cite{Larkin1979pinning}. Neglecting the possibility of disorder-induced domain walls, we expand the double-well potential $((\nabla\phi)^2-Q^2)^2$ in one well as $(\nabla\phi-Q)^2$, which leads to the usual $(\nabla\phi)^2$ term leading to a finite $R$ in $d<4$, and a $Q\nabla\phi$ term (related to so-called nonlocal elasticity \cite{Nattermann2014EPL}), leading to a long-range order already in $d=2$ \cite{NoteM}. This implies that the phason $\phi_i$ in the outer domains will be collectively pinned by impurity potential. Impurities with perpendicular easy axes give rise to an isotropic potential, hence, such impurities must be mapped on a random walk with steps in opposite directions giving total anisotropy strength $\sim k\sqrt{n}$ determining the barrier for the threshold field to overcome, $E_T\sim k\sqrt{n}$, consistent with the exponent $1/4$ in our simulations. This field must also be proportional to $\sqrt{L_i}$, the outer domain width. Such peculiar scaling is the result of the DW motion controlled by phason modes $\phi_i$ and the nonlocal elasticity of spiral magnets.

{\bf Discussion ---} The results suggest that dynamics, dielectric response and disorder effects are very different for type I and type II DWs. 

As seen in Fig.~\ref{fig:type12}(c), type-I DWs are stabilized by a significant easy-plane anisotropy. Such situation may be realized in TbMnO$_3$ with unquenched orbital moment on Mn, where a peculiar polarization ``relaxation'' has been observed under an applied electric field \cite{Biesenkamp2021}: once the field is applied, there is a fast polarization change, while polarization evolution persists at much longer times (hours) although slows down. This behavior is consistent with our predictions for type-I walls, which move quickly once nucleated at the boundary, and slow down substantially inside the bulk, as given by Equations~(\ref{eq:X2},\ref{eq:Xt}). Recent experiments on NaFeGe$_2$O$_6$ indicate the maximal DW velocity comparable with the magnon speed \cite{Biesenkamp2021}. In the data interpretation the velocity is usually assumed to be determined by an external field and independent of the domain structure, and the flattening of the polarization curve is attributed to disorder and wall pinning \cite{Stein2021,Biesenkamp2021}. As we see from Eq.~(\ref{eq:X2}), both high initial velocity and the slowdown can be naturally explained without any reference to disorder: the slowdown is directly related to the increasing number of spins that have to be rotated to move the DWs further. Indeed, in experiments on TbMnO$_3$ \cite{Stein2021} and NaFeGe$_2$O$_6$ \cite{Biesenkamp2021}, the polarization is changing rapidly after the field application, and then slows down dramatically, evolving at kilo-second timescale. 
In real materials complex domain structures are usually present. However, the walls whose plane contains the wave vector, the vortex walls, have a much higher energy (surface tension) compared to the walls perpendicular to the wave vector, and thus they are quickly eliminated when the sample is cooled through the transition into the spiral phase \cite{Li2012}.

{


In summary, in spiral multiferroics magnetic domains are also ferroelectric, and therefore magnetism can be controlled by electric fields. The resulting multiferroic DWs have several configurations: high-energy vortex walls form when the wave vector belongs to the spin rotation plane; for the perpendicular wave vector, DWs come in two types. Weak easy-plane anisotropy leads to type-II walls that behave similar to more conventional walls in (anti)ferromagnets. Stronger anisotropy leads to type-I walls. Here, the spin structure is characterized by the continuous polar angle, with the ferroelectric polarization given by its winding number. This topological nature of type I walls leads to unconventional equations of motion. Only the outermost walls, closest to the surface, move under an external electric field. Their motion is nonlocal, as it requires rotations of all spins at least in one of the adjacent domains (relies on phason modes), and their velocity depends on the domain structure: while they are slow deep inside large samples, for miniature devices the wall velocity grows inversely proportional to the device size and can approach the magnon velocity. 
The dielectric responses from type-I and type-II walls are very different: for the former, only the outermost walls move, and are obstructed by bulk pinning and Pierls-Nabarro barriers, thus giving zero response in the thermodynamic limit. Type II DWs give bulk response, since all the walls move and they are wide and mobile.

These features of type-I DWs may be beneficial for creating multiferroic memory devices, that could harness electric control and fast switching rate of small bits. We expect the work to inspire future research in spiral multiferroics, new applications and technology. Opportunities for future research include the study of evolution of realistic domain structures and systems beyond easy-plane spiral magnets. Possible boundary effects on domain nucleation, breakdown scenarious at higher fields, and effects due to wall proximity must be studied.

\bibliography{biblio}

\widetext
\newpage
\renewcommand{\thetable}{S\arabic{table}}%
\renewcommand{\thefigure}{S\arabic{figure}}
\renewcommand{\theequation}{S.\arabic{equation}}
\setcounter{equation}{0}
\setcounter{figure}{0}
\setcounter{table}{0}

\begin{center}
\textbf{\large SUPPLEMENTARY MATERIAL}
\end{center}
{\bf Energetics of type I and type II domain walls}
\\

In order to minimize the energy from Eq.~\ref{eq:Lag}, 
\begin{equation}
E=\int dx\left(J(\nabla\mathbf{M})^2+J'(\nabla^2\mathbf{M})^2 +K_z M_z^2\right),\label{eqS:ene}
\end{equation}
we introduce $\mathbf{M}=(\cos\phi(x),\sin\phi(x),0)$ and $q(x)=\phi'(x)$, and the energy takes the form
\begin{equation}
    E=\int dx\left(Jq^2+J'[q^4+(\nabla q)^2]\right).\label{eqS:Etype1}
\end{equation}

By analogy, this can be viewed as an action of a particle with a coordinate $q$ moving in time $x$. Then the equations of motion conserve the Hamiltonian,
$$H=-Jq^2+J'[-q^4+(\nabla q)^2],$$
where we kept the kinetic energy and reversed the sign in front of the potential. In the domain $H=-Jq_0^2-J'q_0^4$ with $q_0=\sqrt{-J/2J'}$. Hence, $J'(\nabla q)^2=J'(q^2-q_0^2)^2$ and we find the shape of the type I wall:
$$\phi'(x)=q(x)=q_0 \tanh{q_0 (x-x_0)}, \quad\phi(x)=\log\cosh{q_0 (x-x_0)}.$$
The wall energy is $$E_{I}=\frac{8}{3}J'q_0^3=\frac{8}{3}J'\left(-\frac{J}{2J'}\right)^{3/2},$$ For type-II walls, one can parameterize the rotation of spins by $\theta\approx q_0 x$ in the spherical coordinates $(\theta,\phi)$ while slow rotation of $\phi$ from 0 to $\pi$ across the wall describes the twisting of the spin rotation plane.

First, we consider a uniform twisting $\phi(x)=\alpha x$ with small $\alpha$. From Eq.~\ref{eqS:ene} we find
$$E=\int dx\left( J(q^2-q_0^2)+J'(q^4+(\nabla q)^2-q_0^4)+\alpha^2\left[J\sin^2\phi+6J'q^2\cos^2\phi\right]\right).$$
Assuming slow twisting for a small $\alpha$, we average $\sin^2\phi\to \frac12,\cos^2\phi\to \frac12$ and obtain an additional energy contribution due to twisting $E_{\alpha}=-\int dx J\alpha^2$, which we can generalize for a non-uniform twisting with the spiral plane normal $\mathbf{n}(x)$ as $E_{\alpha}=\int dx (-J)(\nabla\mathbf{n}(z))^2$. 

In the spirit of a variational solution, we substitute $\cos\phi=\tanh x/\lambda$ and $\theta=q_0 x$ into Eq.~\ref{eq:Lag} and integrate the potential energy over $x$ to obtain the wall energy as 
$$E_{II}=J'\left(Q\lambda^{-3} + 4 Q^3\lambda^{-1} + 
 \frac{Kz}{J'} Q \lambda - \left(\pi Q^2 \lambda^{-2} + \frac{Kz}{J'} \pi Q^2 \lambda^2\right) \left[\sinh(\pi Q \lambda)\right]^{-1}\right).$$ Minimizing $E_{II}$ with respect to $\lambda$ and comparing with $E_{II}$, we obtain the phase diagram in Fig.~\ref{fig:type12} (c). Since our variational solution overestimates the energy of the type II wall, the actual type II region is larger.
\\ 
\\
{\bf Frustrated $J_1$-$J_2$ model}
\\

In numerical simulations we employ the discrete $J_1$-$J_2$ model with spin $\mathbf{S_n}$ on site $n$ and competing ferromagnetic nearest neighbor ($J_1$) and antiferromagnetic next nearest neighbor ($J_2$) interactions, 
\begin{equation}
    E=\sum_n \left(J_1 \mathbf{S_n}\cdot\mathbf{S_{n+1}}+J_2 \mathbf{S_n}\cdot\mathbf{S_{n+2}}+K_z (S_n^z)^2\right).\label{eq:j1j2}
\end{equation}
The energy per site for an ideal spiral $S_n=(\cos Qn,\sin Qn, 0)$ takes the form $E=J_1 \cos Q+J_2 \cos 2Q$, representinag a double well potential with minima at $Q=\pm\arccos\left(-\frac{J_1}{4J_2}\right)$. The curvature at the minimum is $E''_{QQ}=4J_2\sin^2Q$. Comparing the wave vector and the curvature with those from the model \ref{eqS:ene}, $q_0=-\frac{J}{2J'}$ and $E''_{QQ}=-4J$, we obtain the relations between the parameters making these models equivalent (and exactly identical for $Q\to 0$):
$$Q=\arccos\left(-\frac{J_1}{4J_2}\right)=-\frac{J}{2J'},\quad E''_{QQ}=4J_2\sin^2Q=-4J.$$

Assuming a spatially uniform twisting of the 
spin rotation plane with a small constant pitch $\alpha$, we find the energy density
$2 \sin^2 Q J_2 (\nabla n)^2$ (here we use $S(n)=(\cos Q n, \sin Q n, 0)$): 
$$ 
 J_1 S(n)
    \hat{R}_{\alpha, (100)} S(n + 1) + 
  J_2 S(n) 
    R_{2 \alpha, (100)} S(n + 2)\approx J_1 \cos Q+ J_2 \cos 2Q + 2J_1 \sin^2 Q \cdot\alpha^2.$$
This suggests that the energy can be expressed via the spin rotation plane normal $\mathbf{n}(r)$ as 
$$\int dx \left(2J_1 \sin^2 Q (\nabla \mathbf{n})^2-K_z n_z^2\right),$$
which is a sine-Gordon model with the wall energy $E=\sqrt{2J_1 \sin^2 Q K_z }$.

Numerically we fitted the energy of the type I wall to be $$E_{II}=1.95 J_2 \sin^2 Q.$$\\
{\bf Model in the continuum limit}
\\

Here we report the detailed derivation of the equations of motion for the single DW case. 
The dynamics of the spiral magnet is described by the Lagrangian density $\mathcal{L}$ and the Rayleigh dissipation functional $\mathcal{R}$
\begin{equation}
\begin{split}
    \mathcal{L}=\alpha \mathbf{A}(\mathbf{M})\cdot \dot{\mathbf{M}}-J(\nabla\mathbf{M})^2-J'(\nabla^2\mathbf{M})^2 +&\gamma \mathbf{[\mathbf{E}\times \mathbf{x}]}\cdot[\mathbf{M}\times\nabla\mathbf{M}]+k_z M_z^2,\\
    \mathcal{R}=& \beta\dot{\mathbf{M}}^2
\end{split}
\end{equation}
The meaning of the symbols has already been discussed in the main text. We now express  the magnetization in spherical coordinates, 
\begin{equation}
\mathbf{M}(x)=\left\{\begin{array}{c}\sin\theta(x)\cos\phi(x)\\\sin\theta(x)\sin\phi(x)\\\cos\theta(x)\end{array}\right.,
\label{eqS:mag}
\end{equation}
the dependence on $x$ of the angles $\theta$ and $\phi$ is, from now on, omitted for simplicity. \\
We use Eq.~(\ref{eqS:mag}) to rewrite the expressions for $\mathcal{L}$ and $\mathcal{R}$, and obtain
\begin{equation}
\begin{split}
    \mathcal{L}=&\alpha\dot\phi\cos{\theta}-J(\theta'^2+\phi'^2\sin^2{\theta})+J'\phi'^4\sin^2\theta+\gamma E \phi'\sin^2{\theta}+k_z\cos^2{\theta},\\
 \mathcal{R}=&\beta[\dot\phi^2\sin^2{\theta}+\dot\theta^2]. 
    \end{split}
\end{equation}

In order to describe the dynamics of the chiral DW we introduce the Ansatz 
\begin{equation}
\begin{split}
    &\phi=\left\{\begin{array}{cc}
    \phi_0-Q x& \mathrm{if}~x<\bar x\\
    \phi_0-Q \bar{x}+Q(x-\bar x)&\mathrm{if}~x\geq \bar x
    \end{array}\right.\ \\
    &\vphantom{a}\\
    &\cos{\theta}=\left\{\begin{array}{c}
    \delta_1\,\qquad \mathrm{if}~x<\bar x\\
    \delta_2\,\qquad \mathrm{if}~x\geq \bar x,
    \end{array}\right.
\end{split}
\end{equation}
we substitute this expression into $\mathcal{L}$ and $\mathcal{R}$ and then integrate in $x$. However we must point out that, in the expression of $\mathcal{L}$, many terms from $(\nabla^2\mathbf{M})^2$ do not contribute due to the particular form of this Ansatz. The full expression for $(\nabla^2\mathbf{M})^2$ has the form 
\begin{equation}
    (\nabla^2 \mathbf{M})^2=\theta''^2+\theta'^4+2\theta'^2\phi'^2+4\theta'^2\phi'^2\sin^2\theta+\phi''^2\sin^2\theta+\phi'^4\sin^2\theta-2\theta''\phi'^2\cos\theta\sin\theta+4\theta'\phi'\phi''\cos\theta\sin\theta,
\end{equation}
and, given our Ansatz, we realize that all terms containing $\theta'$, $\theta''$ as well as $\phi''$, are  zero, and only the term $\phi'^4\sin^2\theta$ contributes.

After the integration we obtain (we continue to refer to the integrated Lagrangian and Rayleigh functions with $\mathcal{L}$ and $\mathcal{R}$ for simplicity)
{ 
\begin{equation}
\begin{split}
\mathcal{L}=&\alpha\left[\dot\phi_0\delta_1\bar{x}-(\dot\phi_0-2Q\dot{\bar{x}})\delta_2(L-\bar{x})\right]-\mathcal{J}Q^2\left[(1-\delta_1^2)\bar{x}+(1-\delta_2^2)(L-\bar{x})\right]+\\&+Q{\gamma E}\left[-(1-\delta_1^2)\bar{x}+(1-\delta_2^2)(L-\bar{x})\right]+k_z(\delta_1^2 \bar{x}+\delta_2^2(L-\bar{x}))\\    
\mathcal{R}=&{\beta}\left[\vphantom{\frac{\dot\delta_1^2}{1-\delta_1^2}}(\dot\phi_0^2+4Q^2\dot{\bar{x}}^2-4Q\dot{\bar{x}}\dot\phi_0)(1-\delta_2^2)(L-\bar{x})+\dot\phi_0^2(1-\delta_1^2)\bar{x}+\frac{\dot\delta_1^2}{1-\delta_1^2}\bar{x}+\frac{\dot\delta_2^2}{1-\delta_2^2}(L-\bar{x})\right],
\end{split}\label{eq:integrated}
\end{equation}
}
with $\mathcal{J}=J+Q^2J'$. In the model with competing nearest and next-nearest exchange interactions $J_1$ and $J_2$, the equivalent ${\mathcal{J}=J_1+2J_2-\frac{J_1^2}{8 J_2}-k_z}$. We note that the integrals from domains are $\sim x$ and $\sim (L-x)$ while the DW terms are neglected since the DWs in our Ansatz are sharp. Introducing DWs of finite width $\lambda\sim Q^{-1}$ results in terms $\sim\lambda$, that are small compared to the former terms when $\bar{x}\gg \lambda$, $L-\bar{x}\gg\lambda$. These expressions hold when domain size exceeds the domain wall width, as discussed below.

We now use Euler-Lagrange equations in order to derive the equation of motion for the DW:
\begin{equation}
    \frac{\partial \mathcal{L}}{\partial \xi}-\frac{\partial}{\partial t}\frac{\partial \mathcal{L}}{\partial \dot\xi}=\frac{\partial \mathcal{R}}{\partial \dot\xi},
\end{equation}
where $\xi=\bar{x},\phi_0,\delta_1,\delta_2$. We first consider the equations for $\delta_1$ and $\delta_2$,
\begin{equation}
\begin{split}
    \alpha\dot\phi_0+2\Gamma_1\delta_1&=\frac{2\beta\dot\delta_1}{1-\delta_1^2}\\
    \alpha(\dot\phi_0-2Q\dot{\bar{x}})+2\Gamma_2\delta_2&=\frac{2\beta\dot\delta_2}{1-\delta_2^2},
\end{split}    
    \label{eq:phidot}
\end{equation}
with $\Gamma_{1,2}=(\mathcal{J}Q^2\pm Q\gamma E+K_z)$, however,  as the electric field $E$ is small with respect to $\mathcal{J}$, we can write $\Gamma_1\simeq\Gamma_2\simeq\Gamma$. These are precession equations: change of $\phi$ describes the magnetization precession around the $z$ axis, while $\Gamma \delta$ represents the $z$-component of the effective field, acting on the magnetization. The form of these equations suggests a solution composed of a transient part with a timescale of $\beta/\Gamma$ and a steady state term. The small $\delta^2$ term in the denominator is neglected and the solutions for $\delta_{1,2}$ are found as
\begin{equation}
    \begin{split}
        &\delta_1(t)=Ae^{\frac{\Gamma}{\beta}t}-\alpha\frac{\dot\phi_0}{2\Gamma}\\
        &\delta_2(t)=Be^{\frac{\Gamma}{\beta}t}-\alpha\frac{\dot\phi_0-2Q\dot{\bar{x}}}{2\Gamma}.
    \end{split}\label{eqS:deltas}
\end{equation}
Here $A$ and $B$ are constants determined by the initial conditions of the problem. Moreover, $J$ being the dominant term in $\mathcal{J}$ and for a spiral with ferromagnetic nearest-neighbor interactions, it results that $J<0$, so $\Gamma<0$ and the solutions for $\delta_{1,2}$ are stable. Here we also suppose that the terms $\dot{\bar{x}}$ and $\dot{\phi_0}$ are constant, or with a possible transient behaviour that decays faster than the one for the $\delta_{1/2}$.

We then write the equations for $\bar{x}$ and $\phi_0$ and substitute the solutions for $\delta_{1,2}$. We neglect the decaying exponential and truncate the equations to the lowest order in $\dot{\bar{x}}$ and $\dot\phi_0$, that is, as we will see in the following, the lowest order in the external field $E$
\begin{equation}
    \begin{split}
    &\beta(4Q\dot{\bar{x}}-2\dot\phi_0)(L-\bar{x})+\gamma E=0\\
    &(2\dot\phi_0-4Q\dot{\bar{x}})(L-\bar{x})+2\dot\phi_0\bar{x}=0.
    \end{split}
\end{equation}
from which we obtain the equations for the speed of the DW and the rate at which $\phi_0$ changes.
\begin{eqnarray}
    \dot\phi_0&=&-\frac{\gamma E}{2\beta\bar{x}},\label{eqS:phi}\\   
    \dot{\bar{x}}&=&-\frac{\gamma E}{4\beta Q}\frac{L}{\bar{x}(L-\bar{x})}.
    \label{eqS:x}
\end{eqnarray}
We observe that $\dot{\bar{x}}$ and $\dot{\phi}_0$ are both linear in $E$, and inversely proportional to the domain widths, rising to infinity as the wall approaches the boundary.  Accounting for the domain wall width, as discussed above, provides the cutoff for the velocity of DWs near the boundary. Integrating Eq.~(\ref{eqS:x}) we obtain the result reported in the main text
\begin{equation}                                                                              %
    x(t)=\sqrt{x^2(t_0)-\frac{\gamma E}{2\beta Q}(t-t_0)}.                                    %
\end{equation}                                                                                %
As for the solutions for $\delta_{1,2}$, substituting Eq.~(\ref{eqS:phi}), (\ref{eqS:x}) into Eq.~(\ref{eqS:deltas}), we obtain 
\begin{equation}                                                                              %
    \delta_1=\frac{\alpha\gamma}{4\Gamma\beta }\frac{E}{X},\quad                       %
    \delta_2=-\frac{\alpha\gamma}{4\Gamma\beta }\frac{E}{L-X},                       %
\end{equation}                                                                                %

The derivation for the two-DW case is analogous. The expressions for $\phi_0$, $x_1$ and $x_2$ are reported in the main text in Eq.~(\ref{eq:X2}). The expressions for $\delta_{1,2,3}$ have the form:
\begin{equation}                                                                              %
    \delta_1=Ae^{\frac{\Gamma}{\beta}t}+\frac{\alpha}{2\Gamma}\frac{\gamma E}{2\beta x_1},    %
    \qquad \delta_2=Be^{\frac{\Gamma}{\beta}t},                                               %
    \qquad\delta_3=Ce^{\frac{\Gamma}{\beta}t}-\frac{\alpha}{2\Gamma}\frac{\gamma E}{2\beta           (L-x_2)},\label{eqS:delta123}                                                             %
\end{equation}                                                                                %
with $A$, $B$ and $C$ constants determined by initial conditions. We note that $\delta_2=0$ in the steady state, implying that spins in the middle domain do not rotate, which is consistent with Fig.~\ref{fig:simul}(b).
\\
\\
{\bf Multi-domain case}
\\

We can consider the general case with $N$ domain wall by generalizing the procedure shown in the main text. We start with a general expression for the in-plane angle $\phi$
\begin{equation}
    \phi(x)=(-1)^i Qx+\phi_i, \qquad\qquad x_i<x<x_{i+1}
\end{equation}

where $Q$ represents the wave-vector of the spiral and $i$ identifies the $i^{th}$ domain in the chain. The change of sign of the $Q$ term at every site takes into account for the presence of  DWs between two adjacent domains. We describe the out-of-plane component of the spins in the $i^{th}$ domain with $\delta_i$. 

Hence, the generalized Lagrangian results
\begin{equation}
    \begin{split}
        \mathcal{L}=&\alpha[\dot{\phi}_0\delta_0x_1+\dot{\phi}_1\delta_1(x_2-x_1)+...+\dot\phi_i\delta_i(x_{i+1}-x_i)+...+\dot\phi_N\delta_N(L-x_N)]\\
        &-\mathcal{J}Q^2[(1-\delta_0^2)x_1+(1-\delta_1^2)(x_2-x_1)+...+(1-\delta_i^2)(x_{i+1}-x_i)+...+(1-\delta_N^2)(L-x_N)]\\
        +&\gamma Q E[-(1-\delta_0^2)x_1+(1-\delta_1^2)(x_2-x_1)-...
        -(-1)^i(1-\delta_i^2)(x_{i+1}-x_1)-...-(-1)^N(1-\delta_N)^2(L-x_N)]\\
        &+k_z[\delta_0^2x_1+\delta_1^2(x_2-x_1)+...+\delta_N^2(L-x_N)]
    \end{split}
\end{equation}

and the generalized Raileigh function

\begin{equation}
\begin{split}
\mathcal{R}=&\beta[\dot\phi_0^2(1-\delta_0^2)x_1+\dot\phi_1^2(1-\delta_1^2)(x_2-x_1)+...+\dot\phi_i^2(1-\delta_i^2)(x_{i+1}-x_i)+...+\dot\phi_N^2(1-\delta_N^2)(L-x_N)]+   \\
&+\beta\left[\frac{\dot\delta^2_0}{1-\delta^2_0}x_1+\frac{\dot\delta^2_1}{1-\delta^2_1}(x_2-x_1)+...+\frac{\dot\delta^2_i}{1-\delta^2_i}(x_{i+1}-x_i)+...+\frac{\dot\delta^2_N}{1-\delta^2_N}(L-x_N)\right]
\end{split}
\end{equation}

The angle $\phi$ must be continuous at the domain interface $x=x_{i+1}$, i.e. $\phi(x)$ must satisfy
\begin{equation}
    \phi(x_{i+1}^+)=\phi(x_{i+1}^-),
\end{equation}
where the $^+$ or $^-$ apex means that we compute the value of $\phi(x)$ in $x_{i+1}$ coming from right or left respectively. The continuity constraint can then be expanded as
\begin{equation}
(-1)^iQx_{i+1}+\phi_i=(-1)^{i+1}Qx_{i+1}+\phi_{i+1},    
\end{equation}
from which we can write 
\begin{equation}
    \phi_{i+1}-\phi_i=2(-1)^iQx_{i+1}
\end{equation}
and
\begin{equation}
    \phi_i=\phi_0-\sum_{j=1}^i(-1)^j2Qx_j.
    \label{Seq:nonLoc}
\end{equation}
We notice that Eq.~(\ref{Seq:nonLoc}) is a non local relation, where the value of $\phi$ in $i^{th}$ domain depends on the values of $\phi$ in all previous domains.

Now we consider a small oscillation around $(x_i^{(0)},\phi_i^{(0)})$ driven by  $E(t)$, (linear response)     
\begin{equation}
    \begin{split}
        \mathcal{L}=&\alpha[\dot{\phi}_0\delta_0x_1^{(0)}+\dot{\phi}_1\delta_1(x_2^{(0)}-x_1^{(0)})+...+\dot\phi_i\delta_i(x_{i+1}^{(0)}-x_i^{(0)})+...+\dot\phi_N\delta_N(L-x_N^{(0)})]\\
        &-\mathcal{J}Q^2[(1-\delta_0^2)x_1^{(0)}+(1-\delta_1^2)(x_2^{(0)}-x_1^{(0)})+...+(1-\delta_i^2)(x_{i+1}^{(0)}-x_i^{(0)})+...+(1-\delta_N^2)(L-x_N^{(0)})]\\
        +&\gamma E(\phi_0-\phi_N)+k_z[\delta_0^2x_1^{(0)}+\delta_1^2(x_2^{(0)}-x_1^{(0)})+...+\delta_N^2(L-x_N^{(0)})]
    \end{split}
    \label{Seq:L}
\end{equation}

\begin{equation}
\begin{split}
\mathcal{R}=&\beta[\dot\phi_0^2x_1^{(0)}+\dot\phi_1^2(x_2^{(0)}-x_1^{(0)})+...+\dot\phi_i^2(x_{i+1}^{(0)}-x_i^{(0)})+...+\dot\phi_N^2(L-x_N^{(0)})]+   \\
&+\beta\left[\dot\delta^2_0x_1^{(0)}+\dot\delta^2_1(x_2^{(0)}-x_1^{(0)})+...+\dot\delta^2_i(x_{i+1}^{(0)}-x_i^{(0)})+...+\dot\delta^2_N(L-x_N^{(0)})\right]
\end{split}
\label{Seq:R}
\end{equation}

Notice that the electric field term got simplified into $\gamma E(\phi_0-\phi_N)$ by using Eq.~(\ref{Seq:nonLoc}). 

We can rewrite Eqs.~(\ref{Seq:L}-\ref{Seq:R}) as
\begin{equation}
    \mathcal{L}=\mathcal{L}_E+\sum_i^N\mathcal{L}_i, \qquad\qquad\mathcal{L}_i=[\alpha\dot\phi_i\delta_i+(\mathcal{J}Q^2+k_z)\delta_i^2]L_i^{(0)}
\end{equation}
with $\mathcal{L}_E=\gamma E(\phi_0-\phi_N)$, and $L_i^{0} $ is the length of the $i^{th}$ domain. 

\begin{equation}
    \mathcal{R}=\sum_{i=1}^N\mathcal{R}_i,  \qquad\qquad \mathcal{R}_i=\beta(\dot\phi_i^2+\dot\delta_0^2)L_0
\end{equation}

Finally, we can compute the polarization in the system by deriving with respect to the external electric field $E$
\begin{equation}
    P=\frac{\partial \mathcal{L}}{\partial E}=\frac{1}{2}\gamma(\phi_0-\phi_N), 
\end{equation}
we see that only $\phi_0$ and $\phi_N$ are affected by $E$, and $P$ is not proportional to $L$
\\
\\
{\bf Atomistic spin dynamics simulations}
\\

Simulations are done 
with classical spin-1 magnetic ions, the Hamiltonian consistent with Eq.~(\ref{eq:Lag}) and with a magnetic spiral ground state. We started with a state with two domains of opposite chiralities separated by the DW at the center. 
We mimic the action of the external electric field through the DM interaction, $H_{DM}=\alpha[\mathrm{r}_{12}\times \mathbf{P}]\cdot[\mathbf{S}_1\times \mathbf{S}_2]$, where $r_{12}$ is a vector connecting sites 1 and 2, $\mathbf{P}$ is a ferroelectric polarization induced by the displacement of an Oxygen ions linking magnetic ions 1 and 2. In a paramagnetic state the system is centrosymmetric and $P=0$. The application of an external electric field leads to $\mathbf{P}=\chi \mathbf{E}$, that produces the DM-like term in the Hamiltonian, $H_{DM}=\mathbf{D}\cdot[\mathbf{S}_1\times \mathbf{S}_2]$ with $\mathbf{D}=\alpha\chi[\mathbf{r}_{12}\times \mathbf E]$. It is this value of $\mathbf{D}$ that we use in UppASD to simulate the external electric field. 

We simulate the dynamics governed by the LLG equation \cite{Landau1935},
\begin{equation}
\frac{d \mathbf{m}_i}{d t}=-\frac{g}{1+a^2}\mathbf{m}_i\times[\mathbf{B}_i+\mathbf{b}_i(t)]
-\frac{g}{m}\frac{a}{1+a^2}\mathbf{m}_i\times(\mathbf{m}_i\times[\mathbf{B}_i+\mathbf{b}_i(t)]),
\label{eqS:LLG}
\end{equation}
where $\textbf{m}_i$ is the magnetic moment at site $i$, with modulus $m$,
$\textbf{m}_i=g\mu_B/\hbar\textbf{S}_i$. 
the coefficient $g$ is the gyromagnetic ratio, $a$ is the damping parameter. $\textbf{B}_i$ is the magnetic field acting on the moment $\textbf{m}_i$, and $\textbf{b}_i(t)$ is a stochastic magnetic field that mimics the effect of finite temperature by applying random torques on spins.


The values of the exchange parameters are $|J|=0.034641~\mathrm{mRy}\approx~0.47\mathrm{meV}$ for nearest neighbors and $|J'|=0.01~\mathrm{mRy}\approx~0.14\mathrm{meV}$ for next-to-nearest neighbors. Simulations have been performed at $T=0~\mathrm{K}$ and $0.05$ damping. To simulate type-II DWs in Fig.~\ref{fig:disorder}(a),(e), a cell of size $800\times 1\times 1$, with periodic boundary conditions along $\textbf x$ and open boundary conditions along $\textbf y$ and $\textbf z$, has been used. The  spin dynamics was simulated with DM-like term $D=0.00015$~meV and hard-axis anisotropy along $k_z=5\times 10^{-5}$; the timestep used is 0.1~ps. Being narrower and less mobile, type-I DWs in Fig.~\ref{fig:disorder}(b),(f) were simulated on a cell of size $200\times 1\times 1$, with $D=0.001$~meV, $k_z=5\times 10^{-3}$ and timestep 0.1~ps. The breakdown scenario shown in Fig.~\ref{fig:disorder}(g) was found for $D=10^{-2}$~meV. In order to simulate disordered systems, we simulated a supercell of $200\times 20$ sites, with ferromagnetic exchange between different chains $|J_f|=5\times 10^{-3}$~mRy, and randomly selected pinning sites; pinning is represented by easy-axis anisotropy, with the axis randomly selected in every pinning site. In Fig.~\ref{fig:disorder}(c) the DM-like term is $D=9.5\times 10^{-4}$, the density of impurities is $1\%$ and the strength of anisotropy is $k=0.01$. 
In Fig.~\ref{fig:disorder}(d) the DM-like term is $D=10^{-3}$~meV, the density of impurities is $3\%$ and the strength of anisotropy is $k=0.019$. Every point in Fig.~\ref{fig:disorder}(h) is obtained by averaging over 10 different disorder configurations. 

The simulation parameters for Fig.~\ref{fig:fitcompare} are $D=0.01,0.02,0.03$~meV%
, damping of 0.005 and T=0 K. As DM interaction is increased, the agreement with the linear-response model worsens, but qualitative agreement remains.
\\
\\


   

\noindent{\bf Comparison of the DW velocities and trajectories from theory and simulations}
\\

We now compare the motion of the two walls with the predictions of Eq.~(\ref{eq:X2}). 
Fig.~\ref{fig:fitcompare} reports the DW trajectories from the simulations (dots) and the behaviour predicted by the model (solid lines).
We find good agreement between the simulation and the theory. Differences near $t=0$ may be due to a transient effect that is beyond our steady state solution. At large $E$-fields the differences between our linear-response theory and the simulations become more visible.

Fig.~\ref{figS:P(t)-long} shows the simulated time-dependent polarization that results from DW motion in Fig.~\ref{fig:simul}. The initial polarization is zero as initially the two DWs with opposite chiralities and therefore opposite polarizations have the same size. When the walls start moving, the domain with polarization along the field grows, and the polarization varies according to the positions of the outermost DWs, $P(t)\propto {x_1}(t)-{x_2}(t)$, until it saturates when only one domain is left. 
\begin{figure}[ht]
    \centering
    \includegraphics[height=0.4\lw]{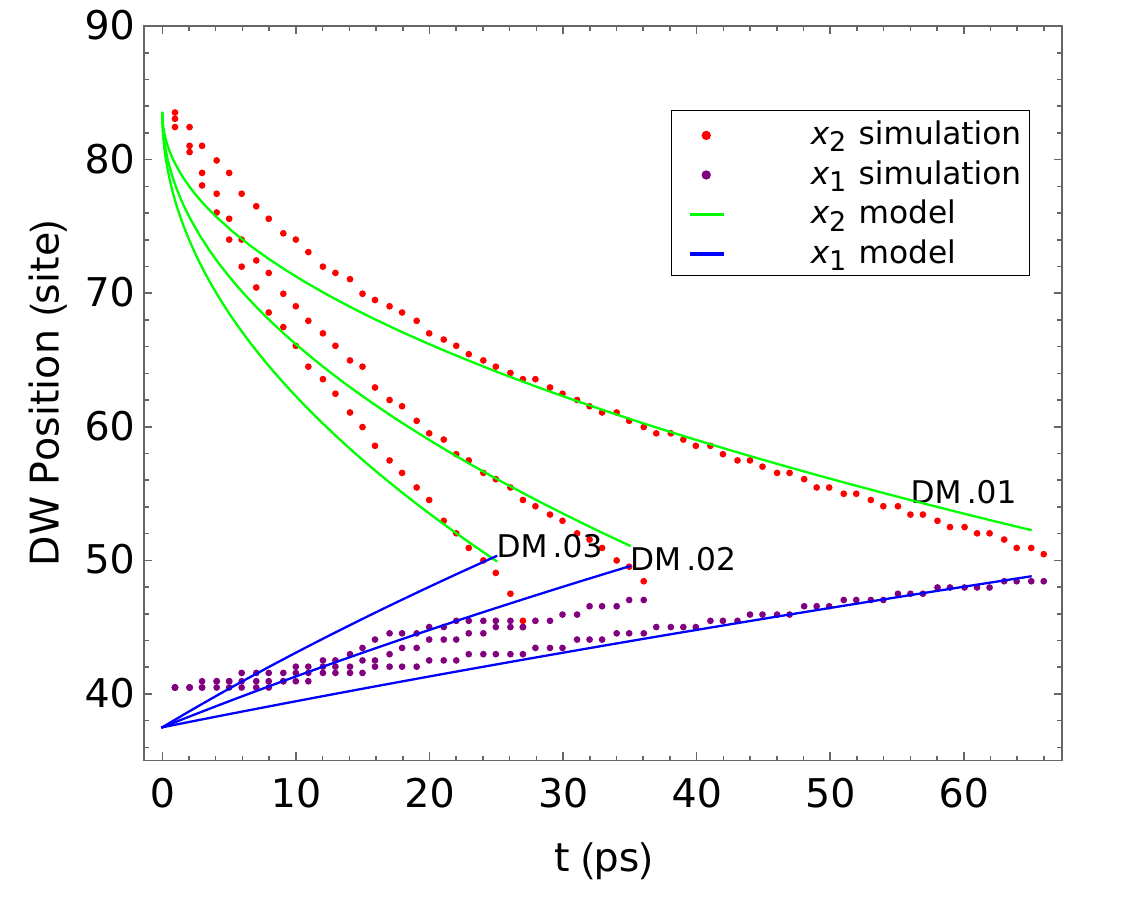}
    \caption{\label{fig:fitcompare} Domain wall trajectories for different field strengths. When the strength of the $E$ field increases, the trajectories deviate from the model derived in the leading order in $E$. Comparison between the analytical model (solid lines) and simulation (dots) of the motion of the two domain walls at positions $x_1$ and $x_2$.\\
    Fitting Eq. $\sqrt{x_{DW}^2+|\gamma E/(2\beta Q)|t}$ (blue line), and $L-\sqrt{|\gamma E/(2\beta Q)|t}$ (green line) 
    }
\end{figure}

\begin{figure}[ht]
    \centering
    \includegraphics[width=.7\lw]{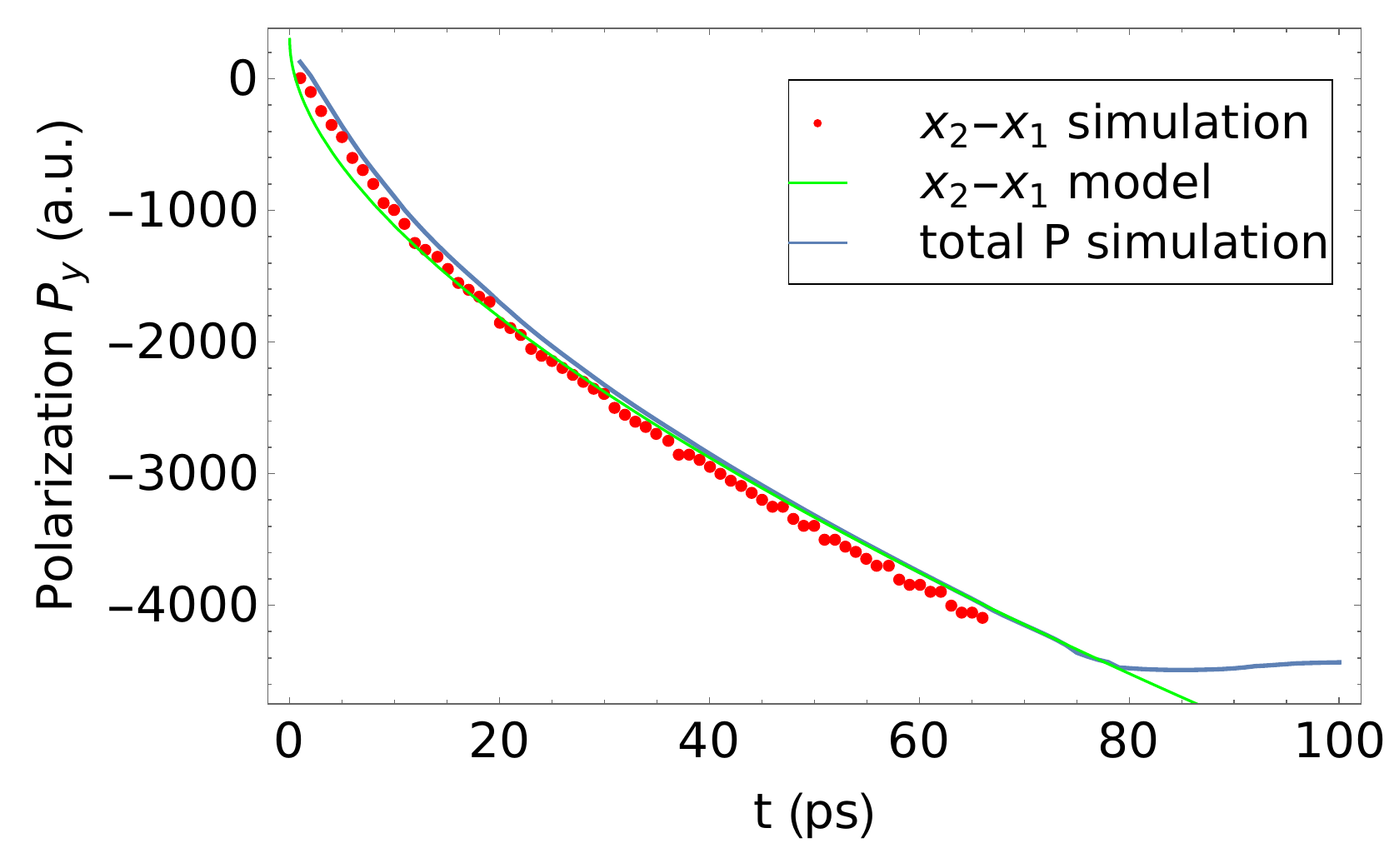}
    \caption{\label{figS:P(t)-long}Time dependence of the polarization. The dependence from the model ($\delta P\sim x_1-x_2$) in green; the same using the DW coordinates from the simulation shown with red dots; the simulated polarization integrated over the sample in blue.}
\end{figure}


\begin{figure}[ht]
    \centering
    \includegraphics[width=0.7\lw]{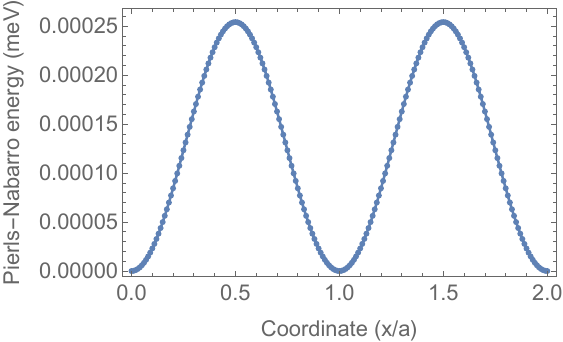}
    \caption{\label{figS:PNbarrier}Pierls-Nabarro barrier for a type I wall with $Q=\pi/4$ and $K_z/J_2=0.1$. Type-II DWs are much wider and the Pierls-Nabarro barrier for them is negligible.}
\end{figure}

\end{document}